\documentclass[doublecol]{epl2} 
\usepackage{graphicx}
\usepackage{amssymb}

\title{Homogeneous nucleation of dislocations as bifurcations in a periodized
discrete elasticity model }
\shorttitle{Homogeneous nucleation of dislocations}
\author{I. Plans\inst{1}, A. Carpio\inst{2} and L. L. Bonilla\inst{1}}
\shortauthor{I. Plans \etal}
\institute{\inst{1} G. Mill\'an Institute for Fluid Dynamics, Nanoscience and Industrial 
Mathematics, Universidad Carlos III de Madrid, 28911 Legan\'es, Spain.\\
\inst{2} Departamento de Matem\'{a}tica Aplicada, Universidad
Complutense de Madrid, 28040 Madrid, Spain}

\pacs{61.72.Bb}{Theories and models of crystal defects}
\pacs{ 05.45.-a}{Nonlinear dynamics and chaos}
\pacs{ 82.40.Bj}{Oscillations, chaos, and bifurcations}

\date{ \today  }

\abstract{
A novel analysis of homogeneous nucleation of dislocations in sheared two-dimensional 
crystals described by periodized discrete elasticity models is presented. When the crystal is 
sheared beyond a critical strain $F=F_{c}$, the strained dislocation-free state becomes unstable via a 
subcritical pitchfork bifurcation. Selecting a fixed final applied strain $F_{f}>F_{c}$, 
different simultaneously stable stationary configurations containing two or four edge 
dislocations may be reached by setting $F=F_{f}t/t_{r}$ during different time intervals 
$t_{r}$. At a characteristic time after $t_{r}$, one or two dipoles are nucleated, split, and the 
resulting two edge dislocations move in opposite directions to the sample boundary. Numerical
continuation shows how configurations with different numbers of edge dislocation pairs 
emerge as bifurcations from the dislocation-free state.}

\begin{document}

\maketitle

\section{Introduction}
Homogeneous nucleation of dislocations is observed in different
processes such as nanoindentation experiments \cite{ase06,rojo1},
heteroepitaxial crystal growth \cite{bre90,joyce}, indentation experiments
in colloidal crystals \cite{sch06} or soap bubble raft models \cite{gou01}. 
Homogeneous nucleation of dislocations occurs in a perfect crystal and is therefore
expected to have a much higher activation energy than heterogeneous nucleation at
defect sites such as step edges. Different types of calculations have been used to interpret 
homogeneous nucleation of dislocations in different situations, ranging from 
atomistic simulations to continuum mechanics interpretations or combinations 
thereof \cite{bulatov}. In all cases, a reliable nucleation criterion is needed to capture
the nature of nucleated defects and the time and place at which such defects appear.

In this letter, we tackle homogeneous nucleation of dislocations as a bifurcation problem 
in discrete elasticity. Periodized discrete elasticity 
models of dislocations regularize in a natural way the singularities of the stress field at the
dislocation lines that appear in the theory of elasticity, and they also allow the sliding
motion of crystal planes typical of dislocation gliding \cite{CB,BCP}. Simplified versions 
of these models have been used to analyze dislocation depinning and motion at the Peierls 
stress in a precise manner \cite{car03}. Here we show that simple periodized
discrete elasticity models
are able to describe homogeneous nucleation of dislocations by shearing an initially 
undisturbed dislocation-free lattice. While molecular dynamics produces nucleation
of dislocations and the Peierls stress, it is considerably costlier than periodized
discrete elasticity.
In more efficient computer codes of discrete line dislocation dynamics, nucleation of 
dislocations and the Peierls stress have to be added to the code \cite{bulatov}. Moreover,
simple models allow for a more detailed analysis and interpretation of results than either
these computer-intensive methods.

The idea behind periodized discrete elasticity models is to discretize space according to 
the nodes of the crystal lattice. Then the potential energy of the crystal is constructed as 
follows. The usual strain energy density of linear elasticity is obtained from the tensor of 
elastic constants and the strain tensor which is the symmetrized gradient of the displacement 
vector. In periodized
discrete elasticity models of simple cubic crystals, the gradient of the displacement vector is 
replaced by a periodic function of the finite differences of the displacement vector that has 
the same period as the lattice and is linear for small differences. The potential energy is 
obtained from the resulting strain energy density by summing over all lattice points 
\cite{CB}. The equations of motion obtained from the crystal potential energy reduce to
those of linear anisotropic elasticity far from the dislocation cores, where the differences of 
the displacement vector are small and approximately equal to their partial derivatives. At the 
dislocation cores, the singularities of the stress field are regularized by the lattice and gliding 
is allowed by the above mentioned periodic functions. Extensions to bcc and fcc
crystals \cite{CB} or to lattices with a two-atom basis \cite{BCP} are possible. While
these models are three-dimensional, it is possible to simplify them for simple geometrical 
configurations such as planar edge dislocations having parallel Burgers vectors. In this case,
the component of the displacement vector parallel to the Burgers vector suffices to describe
depinning and gliding of dislocations. A simplified model retaining only this component has
been used to analyze dislocation depinning and motion at the Peierls stress \cite{car03}. 
The same phenomena occur in the same way in the more complete models 
having two nonzero components of the displacement vector \cite{CB}.

\section{Model}
Here we study a simple 2D periodized
discrete elasticity scalar model that may describe homogeneous 
nucleation of edge dislocations with parallel Burgers vectors and is amenable to detailed 
analysis. Consider a 2D simple cubic lattice with lattice constant normalized to 1, lattice 
points labelled by indices $(i,j)$, $i=1,...,N_x$ and $j=1,...,N_y$, and displacement vector 
$(u_{i,j},0)$. At the boundary, a shear strain $F$ is applied, so that the displacement $u_{i,j}$ 
for $j=1, N_{y}$ and for $i=1, N_{x}$ is $F\, [j-(N_{y}+1)/2]$. $F$ is also the dimensionless 
shear stress. To visualize more easily the lattice, we will sometimes use coordinates $(x,y)$ 
with $x=i-(N_{x}+1)/2$ and $y=j-(N_{y}+1)/2$ centered at the lattice center. In these 
coordinates, the boundary displacement adopts its usual form $Fy$. $u_{i,j}$ obeys the 
following nondimensional equations:
\begin{eqnarray}
&& m {d^2 u_{i,j} \over d t^2} + \alpha {d u_{i,j} 
\over d t} =  u_{i+1,j}-2u_{i,j}+u_{i-1,j}\nonumber\\
&&\quad\quad + A \, [g_{a}(u_{i,j+1}-u_{i,j}) + g_{a}(u_{i,j-1}-u_{i,j}) ]. 
\label{sh1}
\end{eqnarray}
Here $A=C_{44}/C_{11}$ provided we consider cubic crystals with elastic constants $C_{11}$,
$C_{12}$, $C_{44}$. If we select a nondimensional time scale $C_{11}t/(\rho l^2\gamma)
\to t$, then $\alpha=1$ and $m=C_{11}/(\rho l^2\gamma^2)$, where $\gamma$ is a friction 
coefficient with units of frequency, $\rho$ is the mass density and $l$ the dimensional lattice 
constant. The nondimensional displacement vector is measured in units of $l$. With this 
choice of scales, we can consider the overdamped case with $m=0$. On the other hand, if we 
select a nondimensional time scale $C_{11}^{1/2}t/(l\rho^{1/2})\to t$, then $m=1$ and 
$\alpha=l\gamma\sqrt{\rho/C_{11}}$. With this second choice of scales, we can consider 
the conservative case with $\alpha=0$. 

The nonlinear function $g_a$ is periodic, with period equal to the lattice space and $g_a'(0)=
1$. It allows gliding of half columns of atoms in the $x$-direction, which is the direction of 
the Burgers vectors of the edge dislocations that will be nucleated. Gliding along other directions
is not possible in this model: we would need a two-component displacement vector and a
periodic function of finite differences along the $x$ axis \cite{CB}. We have used in our 
simulations the continuous one-parameter family of periodic functions
\begin{eqnarray}
g_a(x)= {2a\over \pi}\left\{\begin{array}{ll}
\sin\left({\pi x\over 2 a}\right), & -a \leq x \leq a,\\
\sin\left({\pi(x-1/2)\over 2a-1}\right), & a \leq x \leq 1-a,\\
\end{array}\right.
\label{sh3}
\end{eqnarray}
with $0\leq a\leq 1/2$ and period 1. In the symmetric case $a=1/4$, (\ref{sh1}) -
(\ref{sh3}) is the interacting atomic chains model \cite{lkk}. The parameter $a$ controls 
the asymmetry of $g_a$, which in turn determines the size of the dislocation core and the 
Peierls stress needed for a dislocation to start moving \cite{CB}. As $a$ increases, the 
interval over which $g'_{a}(x)>0$ increases at the expense of the interval over which
the slope of $g_{a}$ is negative. Then as $a$ increases, so does the Peierls stress, whereas 
both the core size and the mobility of defects decrease. Large values of $a$ result in very 
narrow cores and large Peierls stresses\footnote{Note that the parameter $\alpha$
used in \cite{CB} corresponds to $-a+1/2$ in (\ref{sh3}) and therefore the Peierls stress
in Figure 2 of \cite{CB} decreases as $\alpha$ increases.}. The value of $a$ can be 
selected so that the Peierls stress calculated from (\ref{sh1}) fits experimental values or 
values calculated using molecular dynamics \cite{CB}.

We consider first the overdamped case, $m=0$, $\alpha=1$ with $A=0.3071$ (corresponding
to tungsten), $a=0.2$. In this case and for time independent shear, the potential energy 
\begin{eqnarray}
V = \sum_{i,j}\left[{1\over 2}\,
(u_{i+1,j}-u_{i,j})^2 + A\, G_{a}(u_{i+1,j}-u_{i,j})\right], \label{sh2}
\end{eqnarray}
with $G'_{a}(x)=g_{a}(x)$ and $G_{a}(0)=0$ is a Lyapunov functional of the gradient 
system (\ref{sh1}): it satisfies $V\geq 0$ and 
$${dV\over dt}= \sum_{i,j}{\partial V\over\partial u_{i,j}}\, {d u_{i,j}\over dt} = 
- \sum_{i,j}\left({d u_{i,j}\over dt}\right)^2\leq 0,$$ 
since $du_{i,j}/dt = -\partial V/\partial u_{i,j}$ and the shear strain $F$ does not depend
on time. In the unstressed crystal configuration $F=0$, a given initial condition evolves 
exponentially fast to a stable homogeneous dislocation-free stationary state which we call 
BR0. This stable solution of the discrete model is simply the undisturbed lattice without any 
dislocations. As we select larger and larger positive stresses, the homogeneous stationary 
configuration BR0 is strained but it continues to be stable and dislocation-free until a critical 
stress $F_{c}= 0.2193$ is reached. At $F_{c}$, the maximum eigenvalue of Eq.\ 
(\ref{sh1}) linearized about the stationary solution BR0 becomes zero. What happens for 
$F>F_{c}$?

\begin{figure}
\begin{center}
\includegraphics[width=8cm]{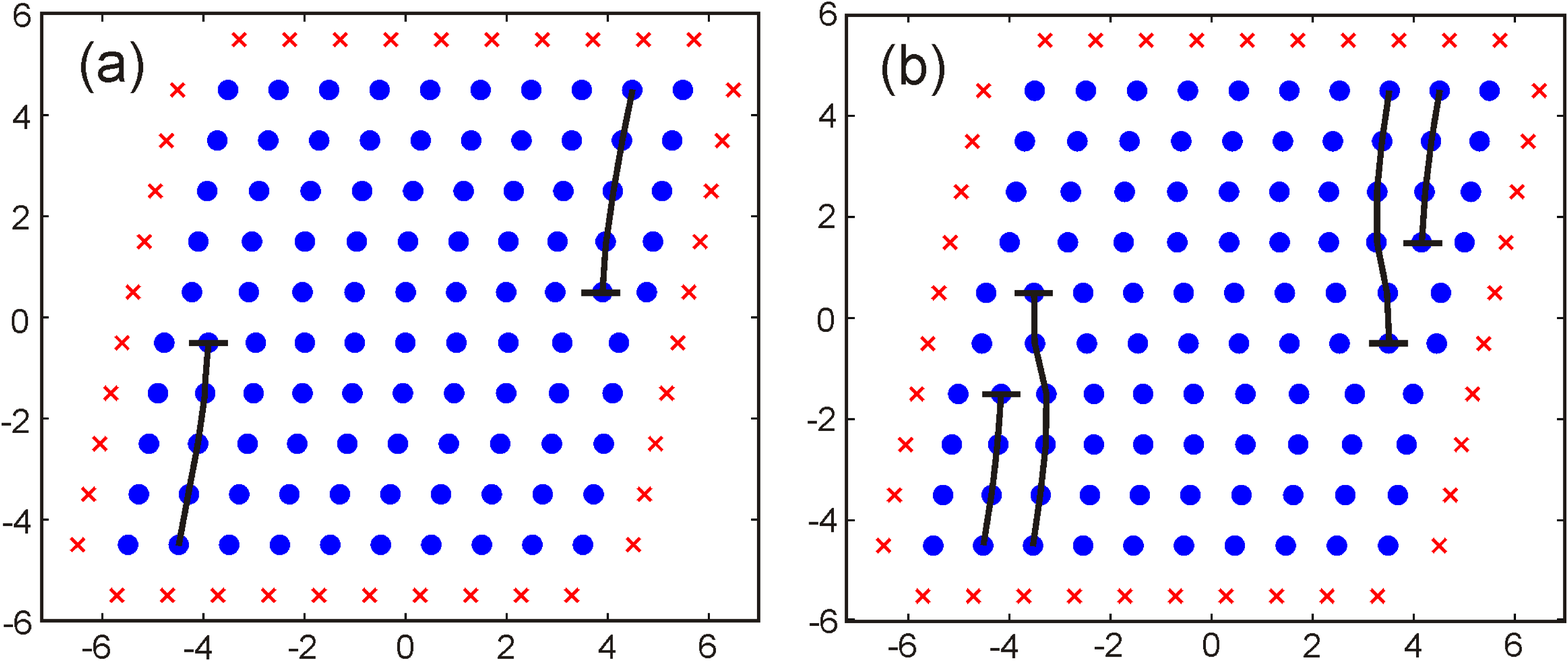}
\caption{Configurations of the stationary solutions (a) BR1, (b) BR2, 
at $F_{f}=0.22$. Parameter values are $a=0.2$, $A=0.3071$ for a 10x10 lattice.
The crosses represents the positions of the boundary atoms which are
fixed by the shear boundary condition.}
\label{fig2}
\end{center}
\end{figure}

\section{Multistable configurations with dislocations} 
One way to proceed is to start from the stable stationary 
configuration BR0 at $F=0$. We then increase the shear strain $F$ 
to a small value $\Delta F$, use the configuration BR0 for $F=0$ as initial condition, solve 
(\ref{sh1}) and find the corresponding stable stationary configuration. Repeating this 
procedure, we follow BR0 until $F_{c}$ and for $F>F_{c}$ we obtain the configuration BR2 
at the corresponding value of $F$ depicted in Fig.~\ref{fig2}(b). The stationary 
configuration contains four edge dislocations, two with Burgers vector $(1,0)$, the other two
with Burgers vectors $(-1,0)$. These dislocations appear as the result of the nucleation of two 
edge dipole dislocations at $y=\pm 1$. Immediately after they are created, these dipoles split 
into their component dislocations that move in opposite directions until they reach the sample 
boundary. Why do dipoles split in opposite moving edge dislocations? The strain needed 
for a given edge dislocation dipole to split into its component dislocations is about $F_{c}/10$
\cite{thesis}, much smaller than the strain $F_{c}$ needed to nucleate a dipole. Thus the
dipoles split into opposite moving dislocations immediately after being created.

 \begin{figure}
\begin{center}
\includegraphics[width=8cm]{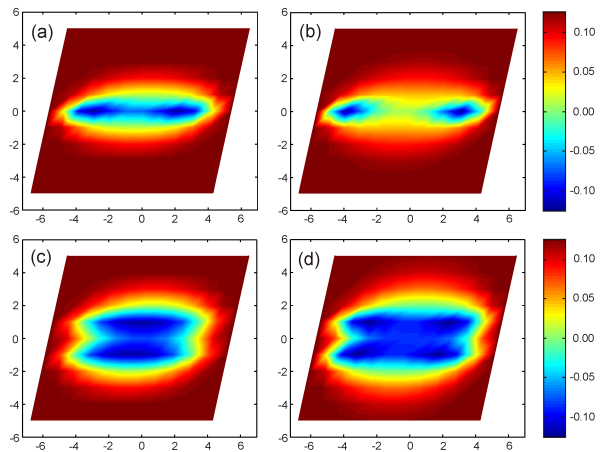}
\caption{Four snapshots of the strain $2 e_{1,2}$ at times (a) 1491.8 and (b)
2643.2 for the evolution towards BR1 at ramping time $t_{r}= 86$ ($c_{r}=0.0026$), and at 
times (c) 2056 and (d) 3242.3 for the evolution towards BR2 at ramping
time $t_{r}=1000$ ($c_{r}=2.2\times 10^{-4}$). $F_{f}=0.22$ and other 
parameter values as in Fig.~\ref{fig2}.}
\label{fig4}
\end{center}
\end{figure}

Can we obtain other configurations by doing things differently? The answer is yes.
Suppose we want to explore the stable stationary configurations at a strain $F_{f}
= 0.22$ slightly larger than $F_{c}$. Starting with BR0 at $F=0$, we turn in the 
strain according to a linear law during a time $t_{r}$ (the ramping time) and leave
$F=F_{f}$ for $t>t_{r}$. Then $F(t)= c_{r}t\, H(t_{r}-t)+ F_{f}H(t-t_{r})$, where $c_{r}
=F_{f}/t_{r}$ and $H(x)$ are the strain rate and the Heaviside unit step function, 
respectively. Same as in other multistable systems \cite{BEA}, 
we obtain different final stable configurations depending on $F_{f}$ and $t_{r}$. For long 
ramping times $t_{r}>87$, we again find BR2. Fig.~\ref{fig4}(c) and (d) show two 
snapshots of the strain component $2e_{1,2}=g_{a}(u_{i,j+1} -u_{i,j})$ taken after $F$ has 
reached its final value $F_{f}$ and $u_{i,j}$ is evolving towards its final stationary 
configuration. We observe two depressions of the strain $e_{1,2}$ at $y=\pm 1$ indicating 
nucleation of two dislocation dipoles. As we have said before, the Peierls stress needed to 
split and move the component dislocations in a dipole is much smaller than the stress required 
for homogeneous nucleation of one dipole. Thus after being nucleated, the edge dislocations 
with opposite Burgers vectors comprising
each dipole immediately move in opposite directions towards the lattice boundaries. The final 
configuration is BR2. For shorter ramping times ($82<t_{r}<87$), Figs.~\ref{fig4}(a) and 
(b) show that only one dislocation dipole is nucleated, splits into two edge dislocations with 
opposite Burgers vectors that then move towards the boundaries in opposite directions. The 
final configuration is BR1 as in Fig.~\ref{fig2}(a). For $t_{r}<81$, a final configuration 
BR3 (similar to that in Fig~\ref{fig2}(b) but not explicitly shown) is reached after two 
dipoles are nucleated at the upper and lower boundaries and their component 
dislocations move to the left and right boundaries in opposite directions.

\begin{figure}
\begin{center}
\includegraphics[width=8cm]{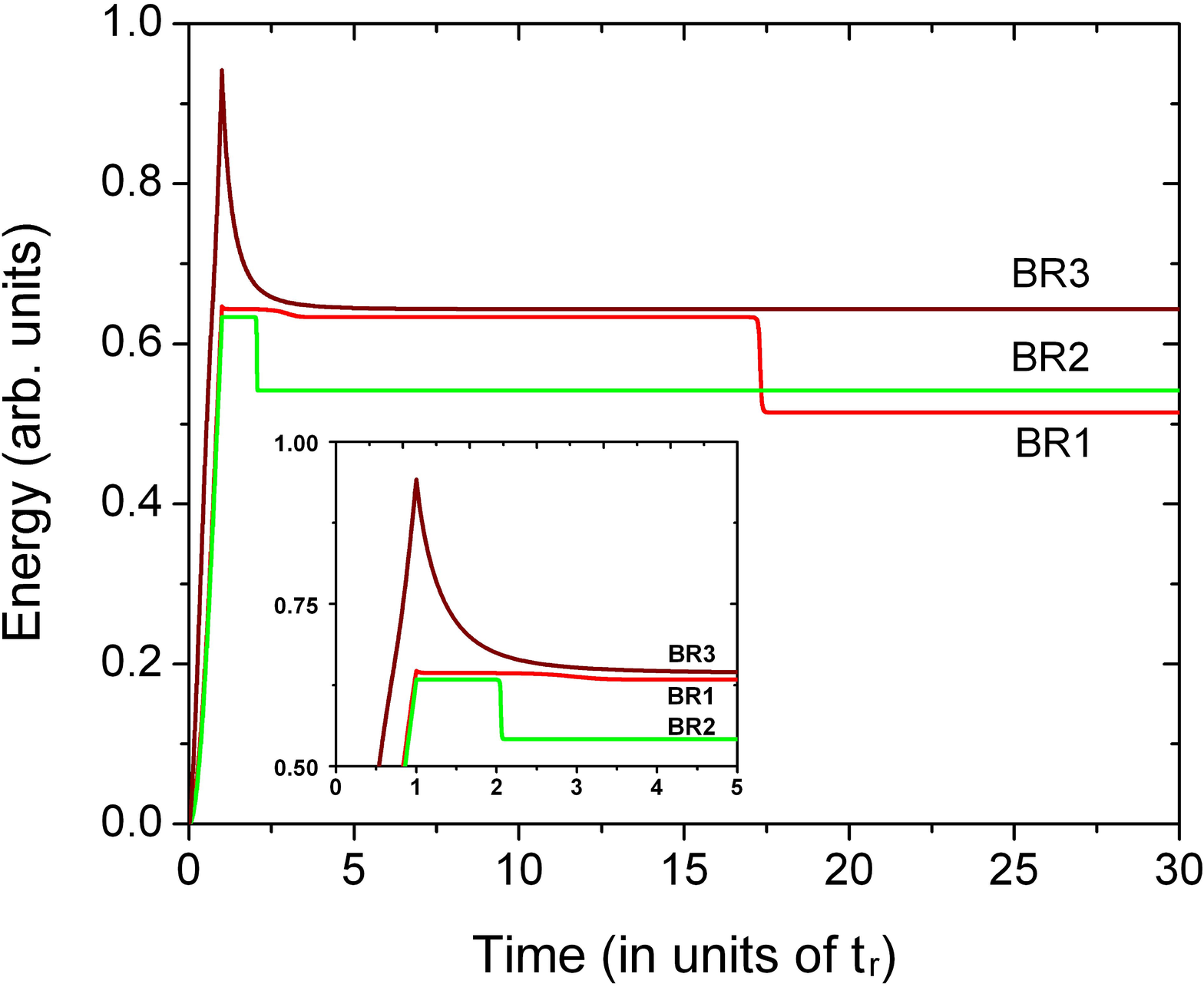}
\caption{Energy $V(t)$ for ramping times 10 (BR3), 86 (BR1) and 1000 (BR2).
Parameter values are the same as in Fig.~\ref{fig2}.}
\label{fig3}
\end{center}
\end{figure}

\section{ Nucleation time} Fig.~\ref{fig3} shows the evolution of the potential 
energy when the same final strain $F_{f}$ is reached at three different strain rates. We
observe that for very short ramping times, e.g., $t_{r}=10$, the energy reaches a peak 
(with strain energy larger than that of the homogeneous branch $V_{\rm BR0}=0.6335$) 
at $t_{r}$ and then relaxes toward its final value $V_{\rm BR3}=0.6433$ corresponding to 
configuration BR3. For very long ramping times, $V(t)$ follows BR0 adiabatically beyond
$t_{r}$, at the plateau $V=V_{\rm BR0}$ that lasts from $t_{r}$ until $2\, t_{r}$. At
this later time, the energy drops abruptly to its final value corresponding to configuration 
BR2 in Fig.~\ref{fig2}(b) with strain energy $V=V_{\rm BR2}=0.5417$. The abrupt 
energy drop marks the nucleation of the two dipoles. Similarly, the precipitous energy drop 
in the evolution towards configuration BR1 of Fig.~\ref{fig2}(a) with energy $V_{\rm 
BR1}=0.5143$ corresponds to the nucleation of one dipole after a long plateau with strain 
energy $V_{\rm BR0}$ ends abruptly at time $17\, t_{r}$. Note that the evolution towards
BR1 starts with a small spike at $t=t_{r}$ (for an intermediate ramping time of 86), it 
continues with energy $V=0.6435$ at a short plateau for times $1<t/t_{r}<2.5$, there is a 
gradual energy decay to the long plateau at $V_{\rm BR0}=0.6335$ that lasts from $t=3.5\, 
t_{r}$ till $t=17\, t_{r}$, and then the strain energy drops to its final value $V_{\rm BR1}=
0.5143$.

\begin{figure}
\begin{center}
\includegraphics[width=8.75cm]{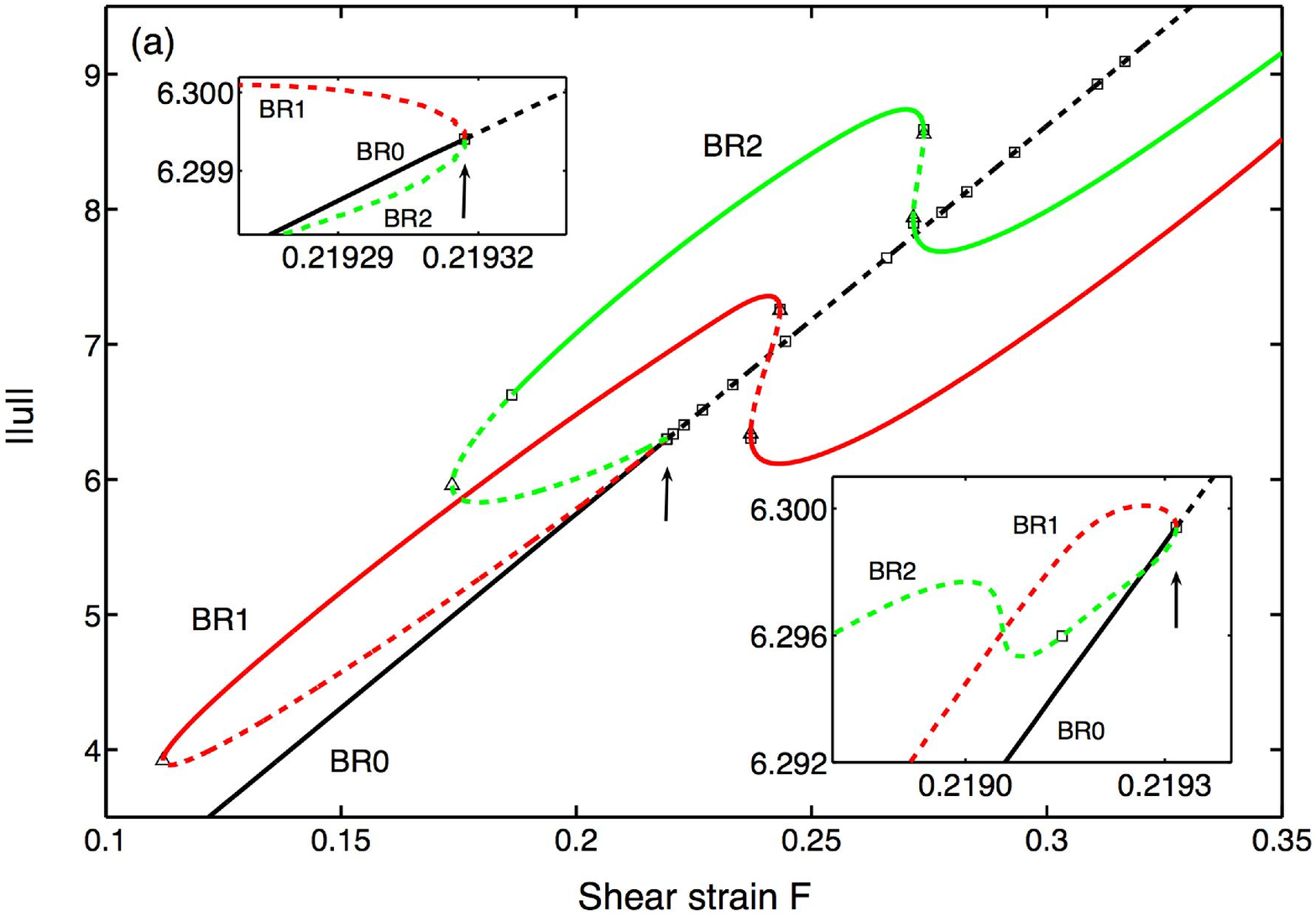}
\includegraphics[width=8.75cm]{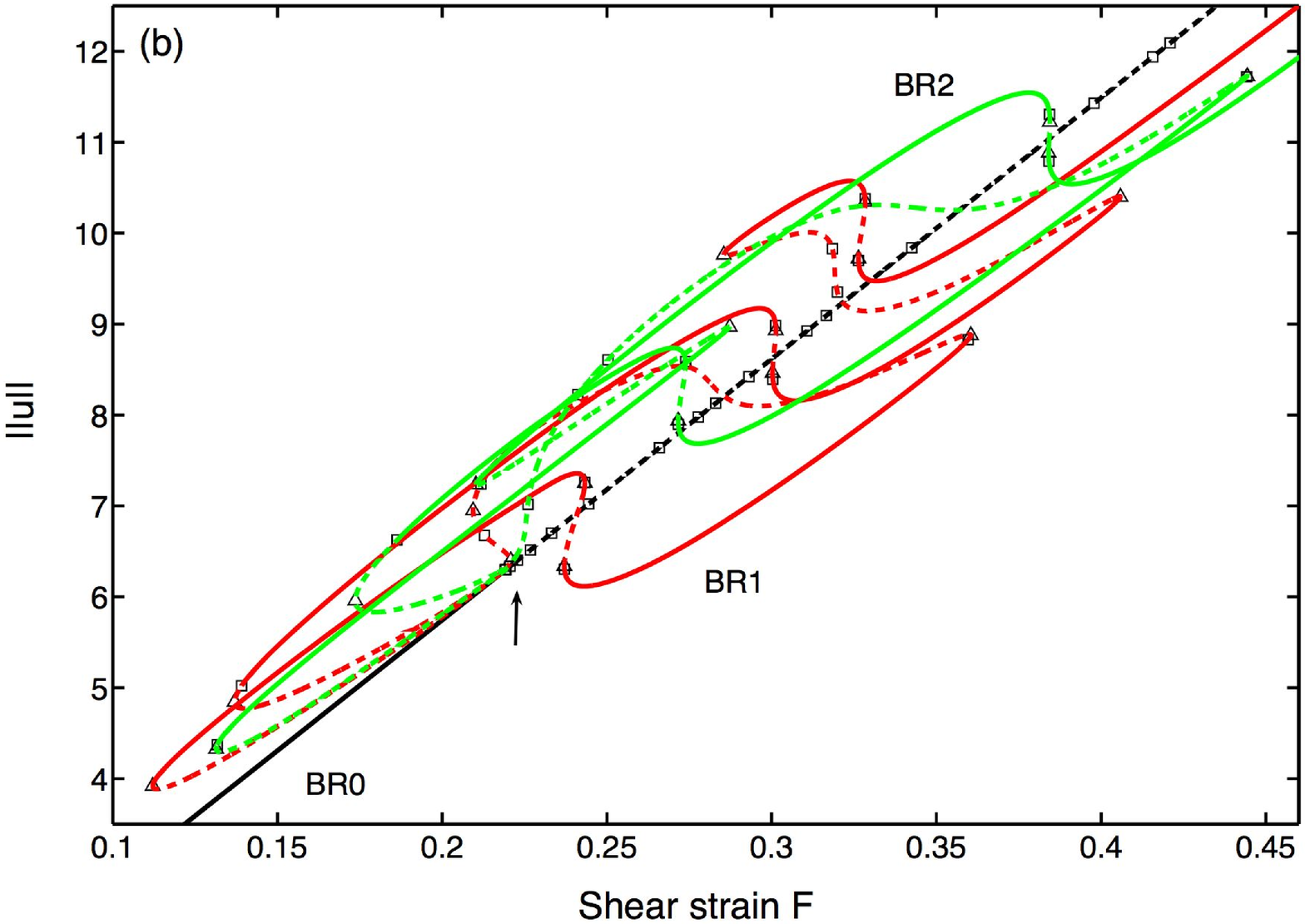}
\caption{(a) Bifurcation diagram showing only the primary stationary branches issuing from 
the homogeneous solution BR0. At $F_{c}$, branches BR1 and BR2 appear as a subcritical 
pitchfork bifurcation from BR0 (see the insets). (b) Larger view of the bifurcation diagram
showing branches that are prolongations of BR1 and BR2 after a number of limit and 
bifurcation points. Solid lines correspond to stable solutions, dashed lines to unstable 
solutions, limit points are marked as triangles and bifurcation points as squares. 
Parameter values are the same as in Fig.~\ref{fig2}.}
\label{fig1}
\end{center}
\end{figure}

\section{Bifurcation diagram} 
More precise information about possible stable configurations can be obtained by examining 
the bifurcation diagram of the $l^2$ norm of the displacement, $||u|| = \sqrt{\sum u_{i,j}^2}$ 
(the sum excludes points at the boundaries), versus the strain $F$.\footnote{We could have
depicted the potential energy $V$ versus strain as a bifurcation diagram. However, the energy 
of the bifurcating branches BR1 and BR2 near the main subcritical bifurcation point is so 
close to the energy of the branch BR0 that visualizing this bifurcation is very hard in an 
energy-strain diagram. Thus we have preferred to use the $l^2$ norm.} The complete 
bifurcation diagram has been calculated using the AUTO program of numerical continuation 
of solutions \cite{auto}, and it is rather complex: there are many bifurcation points issuing 
from different stationary solution branches, most of which are unstable. If we depict all
possible solution branches, the resulting bifurcation diagram is rather messy. Thus we have 
chosen to depict only important solution branches which are stable in certain strain intervals. 

In Fig.~\ref{fig1}(a) we show the only 
two primary branches that bifurcate from BR0 at $F=F_{c}=0.2193$. The inset shows that
these branches appear as a subcritical pitchfork bifurcation at $F=F_{c}$. Both start being 
unstable for $F$ close to $F_{c}$ but become stable after limit points (BR1 exactly after the 
limit point, BR2 becomes stable after a secondary bifurcation point with $F>F_{l2}$), giving 
rise to intervals were several stationary solutions are simultaneously stable. As explained 
before, these configurations can be selected by turning the final shear strain at different rates.

The branches BR1 and BR2 contain a number of secondary bifurcations and
limit points, as depicted in Fig.~\ref{fig1}(b). These branches fold over themselves in 
segments delimited by additional limit points and display other linearly stable parts. The 
configurations thereof contain additional dislocations arising from dipole nucleation. For 
example, BR1 has another stable configuration at $F_{f}$ arising from nucleation 
of dipoles at $y=\pm 2$, whereas BR2 has  stable configurations arising from dipole 
nucleation at $y= 0,\pm 3$ and at $y=\pm 3$, respectively. While these configurations are 
linearly stable, we have not been able to reach them by ramping to $F_{f}=0.22$.

It is interesting to note that the strains $2e_{1,2}$ corresponding to the unstable parts of BR1
and BR2 resemble the snapshots corresponding to dipole nucleation in Fig.~\ref{fig4}.
If we follow the unstable part of BR1 backwards from the limit point at $F_{l1}\approx 
0.11$ to the critical strain $F_{c}$, we observe that its strain $2e_{1,2}$ has a depression
at x = 0 corresponding to dipole nucleation for $F_{l1} < F < F_{c}$, but this depression 
becomes less and less observable as we approach $F_{c}$. Similarly, as we follow the unstable
part of BR2 from its limit point $F_{l2}\approx 0.17$ to $F_{c}$, we observe that $2e_{1,2}$ 
first exhibits two symmetric depressions at x = 0 corresponding to nucleation of two dipoles. 
Then these depressions diminish until the configuration of the unstable part of BR2 becomes 
very similar to that of BR1 as F approaches $F_{c}$. This is as it should be because BR1 and 
BR2 merge at $F_{c}$ in a subcritical pitchfork bifurcation: Near $F_{c}$, BR1 and BR2 
differ from BR0 by $\propto\pm \sqrt{F_{c}-F}\psi_{i,j}$ ($\psi_{i,j}$ is the 
eigenvector corresponding to the zero eigenvalue at $F=F_{c}$).

The components of the eigenvector $\psi_{i,j}$ have opposite signs for alternate rows of 
their corresponding lattice sites\footnote{Note that lattice sites having the same coordinate
$j$ ($y$ axis) form a row and sites having the same coordinate $i$ ($x$ axis) form a column. 
This is different from the usual convention denoting rows and columns of a matrix $A_{i,j}$.
}, $\psi_{i,j}\psi_{i,j+1}<0$, while they keep the same sign along the same column in the
lattice, $\psi_{i,j}\psi_{i,k}>0$. The differences $|\psi_{i,j+1}-\psi_{i,j}|$ are found to 
be largest at the center of the lattice. We have observed that regions where dislocations may 
nucleate are located between rows $j$ and $j+1$ for which $\psi_{i,j+1}-\psi_{i,j}$ is 
maximum ($y=0,\pm 2$ for branch BR1) or minimum ($y=\pm 1,\pm 3$ for branch BR2). 

\section{Influence of lattice size on bifurcations} 
In all our figures, we have presented numerical solutions corresponding to a 
10x10 lattice, $a=0.2$ and $A=0.3071$. The secondary 
solution branches (not depicted here) change substantially with $N_x$, $N_y$, $A$ and $a$. 
Bifurcation and limit points also might appear or disappear from the diagram. However, BR1 
and BR2 persist and the value of $F_{c}$ and the type of the 
primary bifurcation do not change if we increase the computational domain. For example, 
$F_{c}=0.2053$ for a 20x20 lattice which is closer to $F_{c}=a$ than the critical strain for a 
smaller lattice: apparently $F_c \to a$ as the lattice size increases. In 
some cases (small $a$ and small lattices, such as $a=0.1$ in 6x6 and 8x8 lattices), the 
pitchfork bifurcation at $F_{c}$ is supercritical. However, the bifurcation becomes 
again subcritical for larger values of $a$ and for larger lattices ($a=0.1$, in 10x10 and 
14x14 lattices). Finding the branch BR3 with AUTO has been quite elusive and, in fact, we 
did not find it for the 10x10 lattice with the parameter values of Fig.~\ref{fig2}
using AUTO, whereas ramping produced BR3 in a straightforward manner. For a 6x6 lattice 
with $a=0.4$, we find that BR3 appears 
as one of the branches issuing from BR0 as a supercritical pitchfork bifurcation at $F=F_{3}
>F_{c}$. This branch has a limit point at a larger strain and it continues for decreasing 
values of $F$. One of the stretches of BR3 is linearly stable at a range of $F$ overlapping 
those of BR1 and BR2 \cite{thesis}. 

\section{Effects of inertia} The bifurcation diagram corresponding to stationary
solutions of (\ref{sh1}) with inertia ($m\neq 0$) is still the same as presented here. 
However, the stability character of the
solutions changes. In the conservative case $m=1$, $\alpha=0$, stable solutions are
no longer asymptotically stable. Linearizing (\ref{sh1}) about a stable solution, 
we find a problem with purely imaginary eigenvalues. Therefore these solutions
are centers: small disturbances about them give rise to small permanent oscillations
about them. The linearized problem about unstable solutions has pairs of positive
and negative eigenvalues and therefore these solutions are saddle-centers in general.

\section{Concluding remarks}
What have we learned about homogeneous dislocation of nucleations by shearing a 
dislocation-free state? Clearly the {\em critical strain} $F_{c}$ marks the instability of the 
dislocation-free solution branch BR0. $F_{c}$ is characterized as the shear strain at which 
the largest eigenvalue of the linear eigenvalue problem about BR0 becomes zero. The 
components of the corresponding eigenvector indicate possible {\em nucleation sites} that are 
realized by different 
stationary solution branches in the bifurcation diagram. The fact that the pitchfork bifurcation 
at $F_{c}$ is subcritical implies that dislocation nucleation can occur at subcritical shear strain
values at which the solution branches BR1 and BR2 become stable. Thus it is important to 
determine the ranges of $F$ at which some of the branches BR0, BR1 and BR2 are
simultaneously stable. This cannot be done by simple linear stability calculations: instead 
numerical continuation algorithms such as AUTO have to be employed. For overdamped 
dynamics, different stable stationary configurations can be selected by the strain rate at which 
the final strain $F_{f}$ is reached. An abrupt drop in energy marks the {\em nucleation time} 
at which one or two dipoles are nucleated. Since the Peierls stress is lower than the critical 
stress for homogeneous dipole nucleation, the dipoles immediately split into edge dislocations 
with opposite Burgers vectors moving in opposite directions. This bifurcation picture seems to 
describe larger lattices and it captures the stationary solutions even if inertia is added. 

Experimental studies often use ad hoc criteria for nucleation of dislocations such as the 
critical resolved shear stress (CRSS) 
\cite{ase06}. To use this criterion, the critical stress for nucleation has to be related
to the applied force by other means, such as the Hertz contact theory in nanoindentation
experiments \cite{lor03,ase06}. Moreover, the critical stress itself has to be calibrated
independently and it cannot be a fixed value. Instead, the ideal shear stress for nucleation
may depend strongly on the other stress components, not just on the shear stress 
component acting on the plane, as shown by density functional theory \cite{ogata02}.
Earlier continuum mechanics studies suggest that nucleation of dislocations is related to
the loss of strict convexity in the energy and stress concentration \cite{hill}. Recent studies
calculate the elastic constants and internal stresses from atomistic calculations or from
finite element calculations and the Cauchy-Born hypothesis to figure out atom motion 
\cite{zhu04}. Then they minimize a certain scalar functional of elastic constants and
internal stresses at each point of the solid. Nucleation occurs at those points at which the
resulting scalar functional first vanishes \cite{li02}. There is a widespread feeling in all
these studies that homogeneous nucleation of dislocations is related to some bifurcation 
occurring once the instability starts but they do not report any precise analysis
and calculation of this bifurcation, in contrast to our work. 

Dislocation depinning and motion and dislocation interaction occur in the same way in 
the simple scalar model (\ref{sh1}) \cite{car03} and in more complete planar discrete 
elasticity models with two components of the displacement vector \cite{CB}. Thus we 
expect that our bifurcation description of homogeneous nucleation and motion of dislocation 
dipoles also applies to these planar models. Studies of nucleation of dislocations in more 
complete two and three dimensional models are postponed to future publications.

\acknowledgments
We thank L. Truskinovsky for fruitful discussions and J. Gal\'an for help with 
AUTO. This work has been supported by the Spanish Ministry of Education grants 
MAT2005-05730-C02-01 (LLB and IP) and MAT2005-05730-C02-02 (AC), and by 
grants BSCH/UCM PR27/05-13939 and CM/UCM 910143 (AC). I. Plans was financed by 
the Spanish Ministry of Education FPI grant BES-2003-1610.

\end{document}